\documentclass[12pt,preprint]{aastex}
\usepackage{psfig}
\slugcomment{AJ in press}

\shorttitle{CXOCY J125304.0-090737}
\shortauthors{Castander et al.}

\begin{document}

\title{High Redshift X-Ray Selected Quasars:\\ CXOCY J125304.0-090737
joins the club\altaffilmark{1}}

\author{Francisco J Castander$\,$\altaffilmark{2,3,4,5}, Ezequiel
Treister$\,$\altaffilmark{2,3}, Thomas J Maccarone$\,$\altaffilmark{6}, Paolo
S Coppi$\,$\altaffilmark{2}, Jos\'e Maza$\,$\altaffilmark{3}, Stephen
E Zepf$\,$\altaffilmark{7} and Rafael Guzm\'an$\,$\altaffilmark{8}}

\altaffiltext{1}{Partly based on observations collected at the
European Southern Observatory, Chile, under program 68.A-0459}
\altaffiltext{2}{Astronomy Department, Yale University, P.O. Box 208101, New Haven, CT 06520}
\altaffiltext{3}{Departamento de Astronom\'{\i}a, Universidad de
Chile, Casilla 36-D, Santiago, Chile}
\altaffiltext{4}{Andes Prize Fellow}
\altaffiltext{5}{Current address: Institut d'Estudis Espacials de Catalunya/CSIC, Gran Capit\`a 2-4, 08034 Barcelona, Spain}
\altaffiltext{6}{SISSA, via Beirut 2-4, 34014 Trieste, Italy}
\altaffiltext{7}{Department of Physics and Astronomy, Michigan State
University, East Lansing, MI 48824}
\altaffiltext{8}{Department of Astronomy, University of Florida,
P.O. Box 112055, Gainesville, FL 32611}

\begin{abstract}

We present a new X-ray selected high redshift quasar CXOCY
J125304.0-090737 at $z=4.179$, discovered by the Cal\'an-Yale Deep
Extragalactic Research (CYDER) Survey. This quasar is the fifth X-ray
selected high redshift radio quiet quasar ($z>4$) found so far. Here,
we present its observed properties which are characterized by its
relative optical and X-ray faintness, its X-ray hardness and its X-ray
strength compared to optically selected quasars at high redshift. 

We also compare the X-ray selected high redshift radio quiet quasars
to their optically selected counterparts.  We find that the optical
to X-ray spectral slope, $\alpha_{ox}$, is statistically harder (more
X-ray luminous) for the X-ray selected radio quiet quasars than for
the optically selected ones. This result, given the different range of
rest frame ultraviolet luminosities studied and the selection of the
samples, is consistent with the previously found correlation between
X-ray and rest frame ultraviolet luminosities and would extend that
result to a much wider luminosity range at high redshift. Finally, we
discuss the prospects of unveiling the quasar luminosity function at
high redshifts using X-ray surveys.  The discovery of a high redshift
object in the first field of our survey program provides suggestive
evidence that X-ray selected surveys may identify more such objects
than would be expected from an extrapolation of the optical luminosity
function.

\end{abstract}


\section{Introduction}

The formation of the first structures in the Universe is one of the
most interesting subjects in today's astronomy. Because quasars are
intrinsically very luminous, they can be observed up to very high
redshift and can therefore probe early epochs in the universe. Recent
observations have found considerable numbers of quasars at high
redshift \citep{anderson01,schneider02}, with some quasars being at
redshifts $\sim 6$ \citep{fan01b}, when the
Universe was less than one Gyr old.

The earliest quasar surveys were performed in the radio (e.g.,
\citealt{edge59}; \citealt{bennett62}; \citealt{schmidt63}) and the
optical (e.g., \citealt{sandage65}).  While radio surveys have
continued (e.g., \citealt{white00}), in recent years, greater
attention has been placed on optical and X-ray surveys.  Optical
surveys are efficient finding quasars as they can search large areas
of the sky in reasonable amounts of time. At high redshift (in this
paper meaning redshifts $z>4$), the Sloan Digital Sky Survey (SDSS;
\citealt{york00}) has been very successful identifying quasars and now
there are more than 100 SDSS high-redshift quasars known
\citep{anderson01}. However, most of those quasars have been selected
based on their color properties and therefore may be biased against
quasars with different photometric properties like heavily obscured
ones. Recent studies suggest that this bias may not be too problematic
though \citep{richards03}.

The other main quasar selection technique is X-ray detection as X-ray
emission appears to be a universal characteristic of quasars at all
observed redshifts \citep{kaspi00}. However, given the relative
faintness of the majority of the sources, X-ray surveys need to rely
on optical spectroscopy for most of the identifications. The lack of
spatial resolution of past X-ray missions made these efforts laborious
so far. However, the Chandra X-ray satellite with its superb angular
resolution has made the optical identification process much more
efficient. The nature of the X-ray emission also ensures that the
optical selection biases due to obscuration are strongly reduced
(although the X-ray selection may introduce biases of its own). This
is especially true at high redshift where the observed X-ray photons
were emitted at higher energies and can penetrate even
considerably large amounts of obscuring material.

At high redshifts ($z>4$), there are only five quasars known selected
by their X-ray emission. Three of them were found with the ROSAT
satellite \citep{henry94,zickgraf97,schneider98}. The other two were
detected by Chandra \citep{silverman02,barger02}. All of them are
radio quiet except one \citep{zickgraf97}. There is another active
galactic nucleus detected at $z=4.14$ in the Chandra Deep Field North
\citep{barger02}. However, this source is optically faint,
$M_B=-21.7+5\times log (h_{65})$, and therefore it is hard to estimate
the contribution of the active nucleus itself and its putative host
galaxy to the observed optical luminosity. We therefore exclude this
object from further discussions. There is also an optically faint
radio loud active galactic nucleus at $z=4.42$ in the Chandra Deep
Field North \citep{brandt01,vignali02c}, but this object was selected
in the radio. In total there are 56 active galactic
nuclei\footnote{Here, AGN are defined in a broad sense including
quasars and active galaxies} detected in X-rays at $z>4$,\footnote{See
the excellent web page
http://www.astro.psu.edu/users/niel/papers/highz-xray-detected.dat
maintained by Niel Brandt and Christian Vignali for a list of the high
redshift AGN detected in X-rays so far.} seven of which are radio-loud
quasars, two of which are low-luminosity AGN (one of these is
radio-loud) and twelve of which have not appeared in the literature so
far. The remaining 36 sources are radio-quiet quasars and will be the
sample we will discuss in this paper. Thirty-two were optically
selected and four X-ray selected. Twenty-six were detected by Chandra,
nine by ROSAT and one by XMM-Newton. These 36 sources have typical
absorption corrected X-ray fluxes between $10^{-15}$ and
$3\times10^{-14}$ erg cm$^{-2}$ s$^{-1}$ in the 0.5-2.0 keV band.  The
absolute magnitudes of the optically selected quasars range from
$M_B\sim -26$ to $-30 + 5\times log (h_{65})$ while the X-ray selected
quasars range from $M_B\sim -23$ to $-27 + 5\times log (h_{65})$.
Thus, the high redshift X-ray selected quasars sample a different
range of the optical luminosity function than the optical selected
ones.\footnote{Whenever we talk about optical
magnitudes or luminosities we do it in a broad sense including optical
and ultraviolet. The magnitudes in the observed frame are measured in
the optical, however they correspond to rest frame ultraviolet
absolute magnitudes or luminosities.}

There have been several studies of high redshift optically selected
quasars detected in X-rays (e.g.,
\citealt{kaspi00,vignali01,brandt02a,brandt02b,bechtold02,vignali02b,vignali02c}). The
overall properties of the high redshift population are similar to the
low redshift quasars \citep{kaspi00,vignali02b}. The X-ray spectral
slope appears to be similar $\Gamma\sim2.0$ at low and high redshift
(e.g., \citealt{vignali02a,vignali02b}), although in some high
redshift cases it is harder \citep{bechtold02}. The optical-to-X-ray
spectral index, $\alpha_{ox}$, nonetheless, appears to be steeper
(optically stronger and/or X-ray weaker) at high redshift than locally
\citep{vignali01,bechtold02}, although this trend may only be a result
of the possible correlation of the optical-to-X-ray spectral index
with optical luminosity. \cite{vignali02d}, studying a sample of
SDSS Early Data Release quasars, find that $\alpha_{ox}$ is correlated
with rest-frame ultraviolet (UV) luminosity but does not depend
significantly on redshift. They also find a correlation (with a slope
different than unity) between X-ray and rest frame UV luminosities (see
also references therein).

Here we present the discovery of CXOCY J125304.0-090737, a $z=4.179$
quasar selected in X-rays from Chandra archival images. Optical
follow-up images were taken with the CTIO/4-meter telescope and the
confirming optical spectrum was obtained with FORS2 on VLT. These
observations were taken as part of the Cal\'an-Yale Deep Extragalactic
Research (CYDER) Survey \citep{castander03, treister03}. The CYDER
survey is a deep optical and near-infrared imaging and spectroscopic
program encompassing several scientific goals. One of the key aims is
the characterization of the population of faint X-ray sources.
Therefore, some of the fields of the CYDER survey were selected to
overlap with deep Chandra pointings.

In \S 2 we describe the X-ray, optical and near-infrared observations
that lead to the discovery of CXOCY J125304.0-090737, presenting the
data reduction procedures followed. In \S 3 we discuss the observed
characteristics of CXOCY J125304.0-090737 and compare them to those
of the high redshift population of radio quiet quasars. We also
discuss the differences between the X-ray and optically selected high
redshift QSOs. Finally we present our conclusions in \S 4.

Throughout, we assume $H_o=65h_{65}$ km s$^{-1}$ Mpc$^{-1}$,
$\Omega_o=0.3$ and $\Omega_{\Lambda}=0.7$. We define the photon index
$\Gamma$ as the exponent giving a photon flux density X-ray spectrum
$f(E)\propto E^{-\Gamma}$ in photons cm$^{-2}$ s$^{-1}$ keV$^{-1}$.

\section{Observations and Data Analysis}

\subsection{X-ray Data}

The first X-ray field studied by the CYDER survey in the Southern
Hemisphere was the Chandra pointing towards the Hickson group of
galaxies HCG62 centered on (J2000) $\alpha =12^h53^m05.70^s$ $\delta
=-09^o12'20.0''$ (PI: Vrtilek). This field was observed on January 25,
2000 with the ACIS-S detector configuration for a total 49.15 ks. We
retrieved this image from the archive and analyzed it using standard
techniques with the CIAO package.

We remove bad columns and pixels using the guidelines specified on the
ACIS Recipes,\footnote{See Clean the Data at
http://www.astro.psu.edu/xray/acis/recipes/clean.html}, and we remove
flaring pixels using the FLAGFLARE routine.  We use the full set of
standard event grades (0,2,3,4,6) and create two images, one from 0.5
to 2.0 keV and one from 2.0 to 8.0 keV.  We then use the WAVDETECT
routine from the CIAO package to identify the point sources within
these images, checking all wavelet scales from 1 to 16 separated by
factors of 2.  We detect 30 sources from 0.5-2.0 keV and 19 from
2.0-8.0 keV (with 15 detected in both bands, so that a total of 34
sources are detected) in the S3 ACIS CCD.  We then use the PSEXTRACT
script to extract spectra for these sources and provide a response
matrix. CXOCY J125304.0-090737 was detected in both
bands. Figure~\ref{fig2} shows a 30''$\times$30'' cutout region
centered on CXOCY J125304.0-090737 of the total (0.5-8.0 keV) band
adaptively smoothed X-ray image. The total number of background
subtracted photons detected were $17.1\pm4.2$ in the soft [0.5-2.0]
keV band and $8.3\pm3.0$ in the hard [2.0-8.0] keV band.  This
analysis will focus on CXOCY J125304.0-090737; the sample as a whole
will be discussed in a future work.

\begin{figure}[th]
\centerline{\psfig{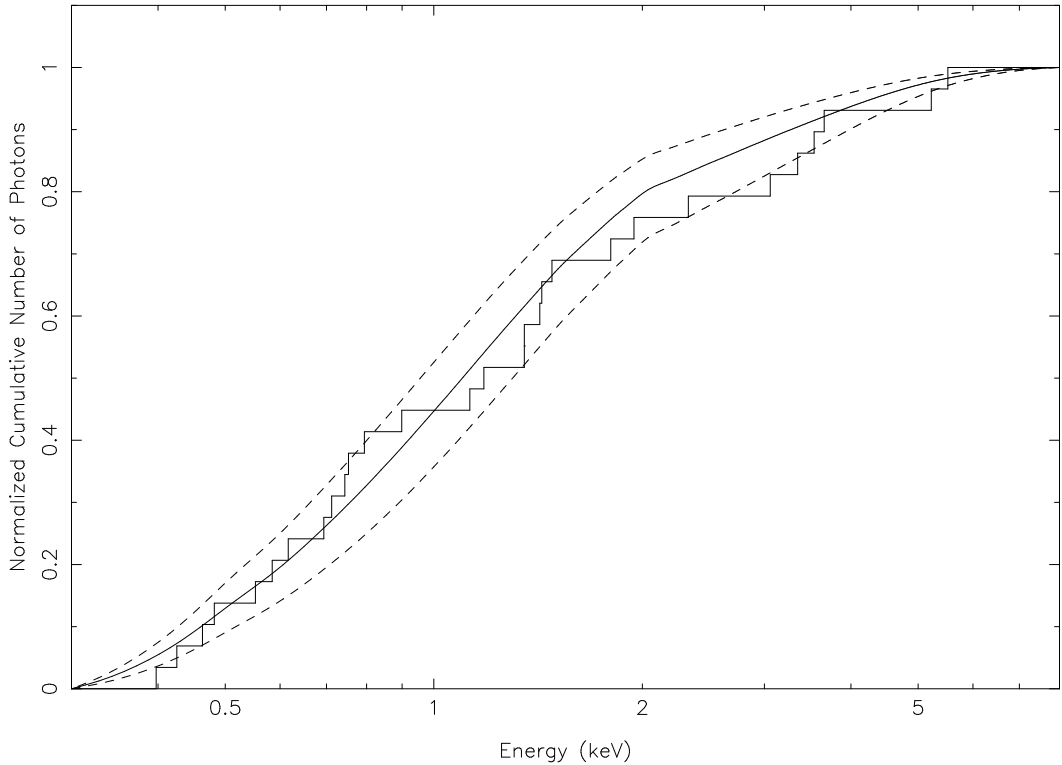}}
\caption{Observed cumulative X-Ray photon energy distribution. The
best fitting power-law model is over-plotted as a solid line, together
with the 1 sigma deviations (dashed lines).\label{fig1}}
\end{figure}

We then fit the spectrum for CXOCY J125304.0-090737 in XSPEC 11.0 with
a power law model.  We find a best fitting photon power law index of
$1.57^{+0.39}_{-0.33}$ with no absorption and of $1.69^{+0.40}_{-0.35}$
($1\sigma$ errors for the spectral indices) when the Galactic neutral
hydrogen column of $3\times10^{20}$ (as calculated using the HEASARC
nH tool) is accounted for.  These fits are done with three spectral
channels from 0.5 to 8.0 keV, which all suffer from statistics too
poor for conventional $\chi^2$ fitting and the resulting values of
$\chi^2$ are close to zero, indicating that the bins have too few
photons.  Still the same results are obtained when either the C
statistic or the $\chi^2$ statistic are used, so they can be
considered as robust as any spectral fits based on less than 30
photons.  The 0.5-8.0 keV Galactic absorption corrected observed flux
is $4.9\pm{1.4}\times10^{-15}$ ergs/sec/cm$^2$ with either model.
Figure~\ref{fig1} shows the observed cumulative energy distribution.

\subsection{Optical and Near-Infrared Imaging Data}

\begin{figure}[th]
\centerline{\psfig{file=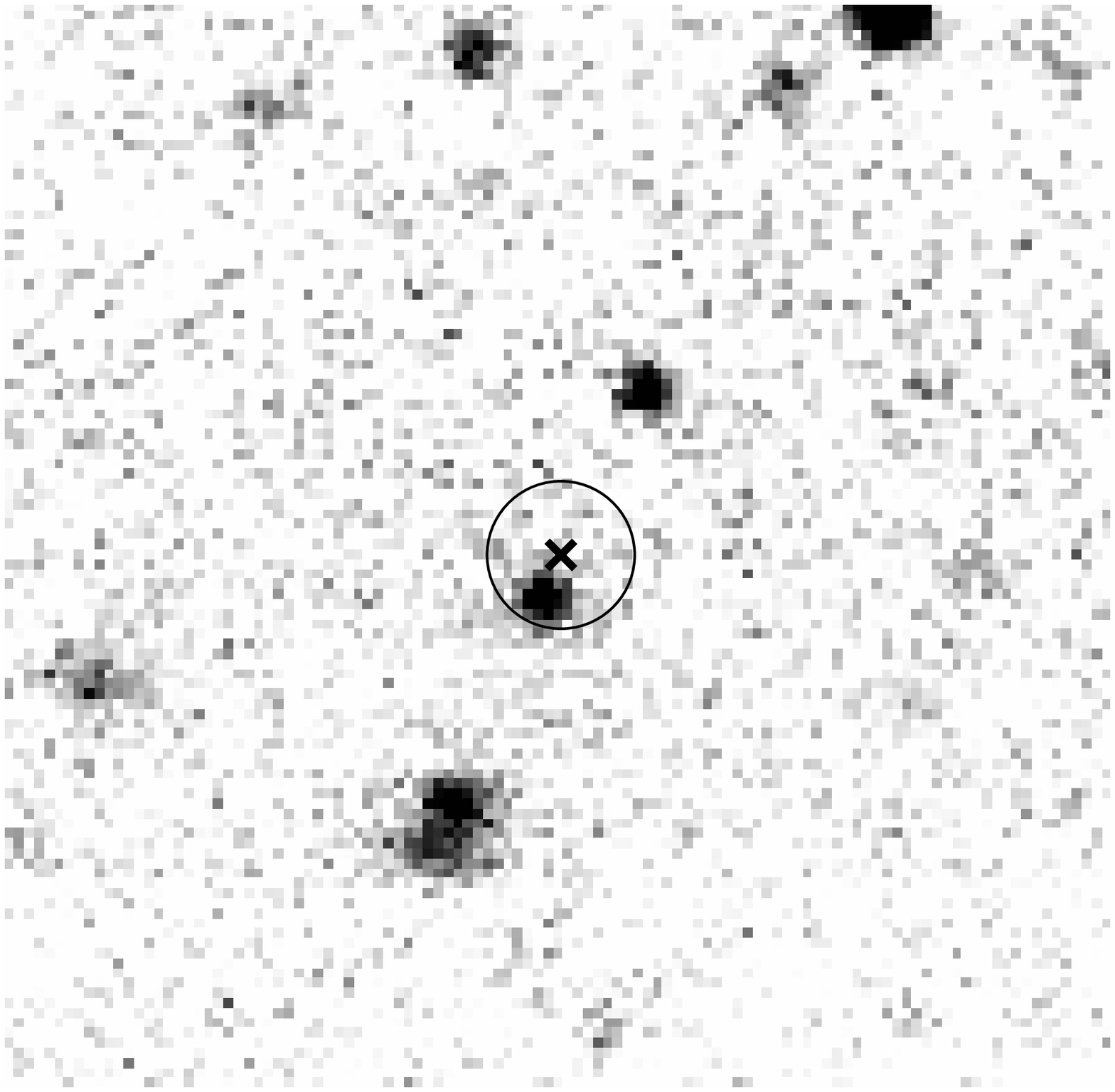,width=7.0cm}\psfig{file=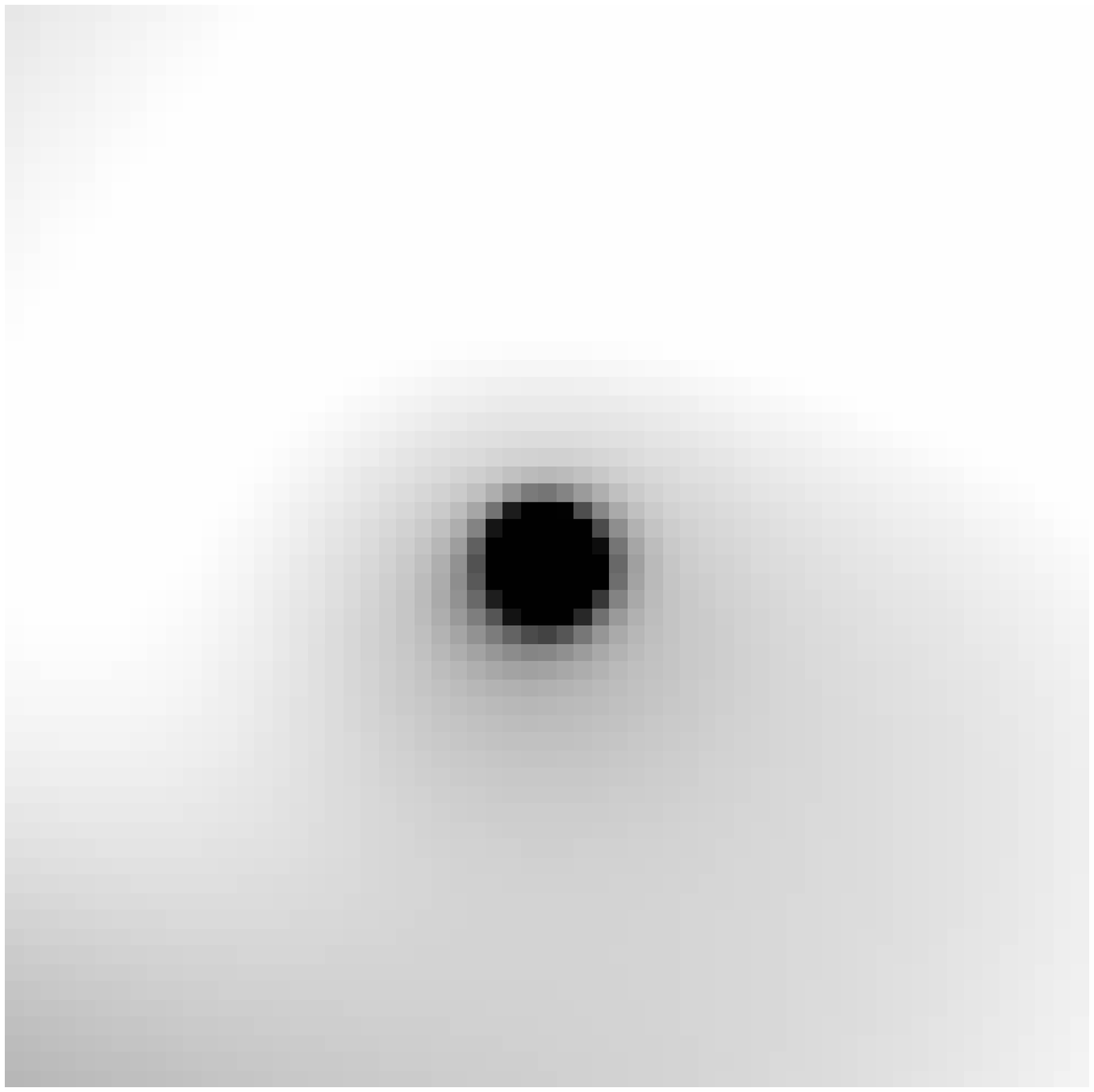,width=7.0cm}}
\caption{V and X-ray images of CXOCY J125304.0-090737. Both images are
cut-outs of 30'' x 30'' of the original images centered on the RA and
DEC of the X-ray source. North is up and East to the left in both
images. The V image (left) was taken at the CTIO-4m mosaic in a total
exposure of 1h 20m. The cross indicates the position of the X-ray
centroid. A 2'' radius circle is also plotted. The X-ray image (right)
was taken with the Chandra X-ray satellite with the ACIS-S chip s3,
exposure time 49 ks. The X-ray image correspond to the full (0.5-8
keV) band and has been adaptively smoothed using the csmooth routine
within CIAO.  We apply a Gaussian kernel to the data, and use a fft
convolution.  The resulting image pixels are required to have a
minimum significance of 3 $\sigma$ and a maximum significance of 5
$\sigma$.\label{fig2}}
\end{figure}

As part of the CYDER survey we obtained V and I images of this field
with the CTIO 4m MOSAIC-II camera.  The total integration time was 80
minutes in V and 25 minutes in I under 1.1'' seeing conditions. Images
were reduced using standard techniques with the IRAF/MSCRED
package.\footnote{IRAF is distributed by the National Optical
Astronomy Observatory, which is operated by the Association of
Universities for Research in Astronomy, Inc., under cooperative
agreement with the National Science Foundation.}  We search for
optical counterparts of our X-ray detections in the resulting
images. The source matching is relative simple given the good spatial
resolution of Chandra. Using the brightest nine point sources in the
optical we find an overall offset between the X-rays and the optical
(tied to the USNO reference frame) of $\Delta RA = 0.9''$ and $\Delta
DEC = 1.0''$ (see Castander et al, in preparation for further
details). There is only one detectable optical counterpart within the
error circle of the X-ray source CXOCY J125304.0-090737 down to our
magnitude limit ($V<25.8$ and $I<23.8$ at 5$\sigma$). The offset
between the X-ray and optical centroids are 1.3''. For this object we
performed PSF fitting photometry measuring $V=23.65\pm0.05$ and
$I=22.52\pm0.08$. Image cut-outs of this object are shown in
Fig~\ref{fig2}.

Near-infrared data for this field was obtained at LCO using the DuPont
2.5m telescope with the Wide Field InfraRed Camera. The total
integration time was 62 minutes in the $J$ band and 60 minutes in the
$K_s$ band, with seeing of 0.7'' and 0.6'' respectively. We reduced
the data with the IRAF DIMSUM package following standard
procedures. We barely detect the optical counterpart in the $J$ band
image at a magnitude of $J=21.90\pm0.20$ but we do not detect it in
the $K_s$ image which reached a magnitude limit of $K_s<19.95$
(5$\sigma$).

\subsection{Optical Spectroscopy}

Follow up spectroscopy of this field was obtained at the ESO Cerro
Paranal Observatory with the UT4/Yepun telescope using the FORS2/MXU
instrument on the 14th February 2002. Three masks were design for the
HCG62 field, one of which included a slit for the CXOCY J125304.0-090737
optical counterpart. We took three exposures with this mask for a
total integration time of 1 hour 55 minutes in $\sim$ 0.9'' seeing
conditions. We used the V300 grism giving a resolution of $R\sim520$
(10.5 {\AA}) for our 1'' slit. The spectrum was reduced using standard
techniques with IRAF and calibrated in wavelength using He-Ar
comparison lamps exposures and the night sky lines. We also obtained
spectra of the spectrophotometric standards Feige67 and LTT6248, which
were used to provide a rough flux calibration of the spectrum.

\begin{figure}[thb]
\centerline{\psfig{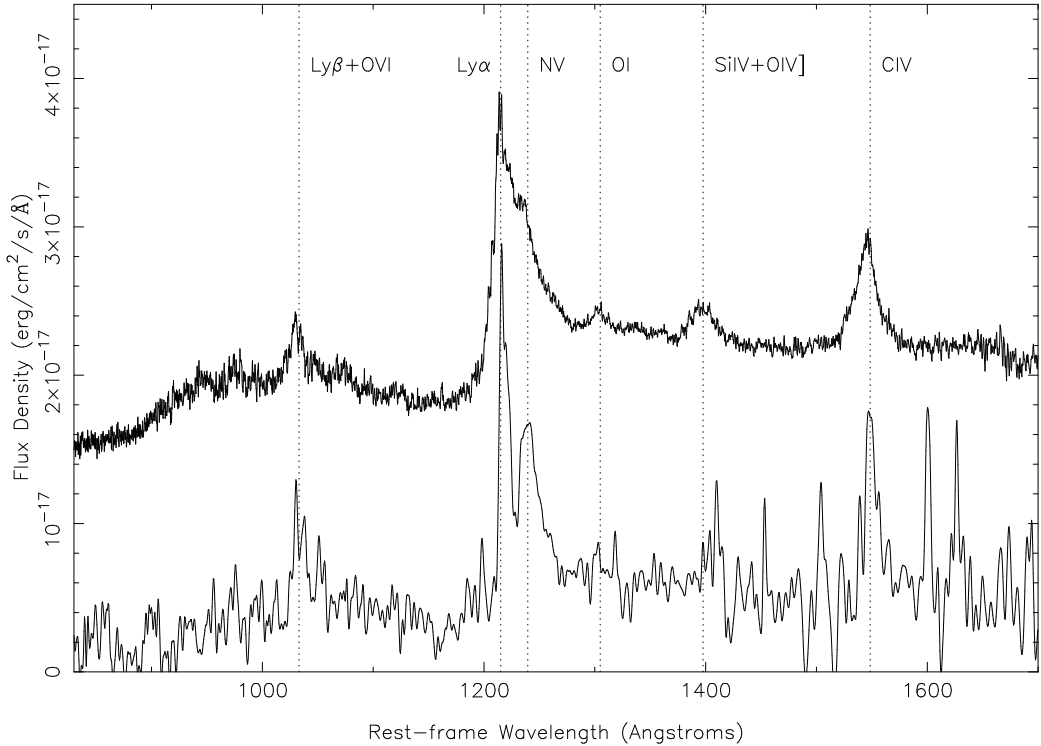}}
\caption{Optical spectrum of CXOCY J125304.0-090737 obtained at
UT4/Yepun VLT. The spectrum has been smoothed in logarithmic
space with a Gaussian of 250 km/s, equal to the spectral
resolution of the instrument for the set up used, $R\sim520$
($v\sim250$ km/s). For comparison, we also show the error weighted
average of the SDSS Early Data Release QSO spectra at $z>4$. Each
individual spectrum used in the $z>4$ composite spectrum has a
resolution of $R\sim1800$ ($v\sim70$ km/s). This composite spectrum
has arbitrary zero point and scaling offsets for display purposes and
it is shown above the spectrum of CXOCY J125304.0-090737. The most
common QSO emission lines are indicated as dotted lines. The emission
lines in CXOCY J125304.0-090737 are narrower than the typical SDSS
spectrum. \label{fig3}}
\end{figure}

Figure~\ref{fig3} shows the final spectrum together with the mean
spectrum of quasars at $z>4$ found in the SDSS. The typical broad
emission lines of a quasar are clearly distinguishable and marked in
the figure. We measure a mean redshift of $z=4.179\pm0.006$ using all
the emission lines but Lyman $\alpha$ and NV.

\section{Discussion}

Table~\ref{tab1} presents the observed properties of CXOCY
J125304.0-090737. In the optical, CXOCY J125304.0-090737 is a faint
quasar with an absolute magnitude, $M_{1450(1+z)}=-23.31+5\times log
(h_{65})$ and $M_B=-23.69+5\times log (h_{65})$. At this absolute
magnitude, one would expect that most of the observed flux is coming
from the QSO itself with little contribution from its host galaxy,
which is not resolved in our optical images. Its spectrum is typical
of a quasar with broad emission lines (Figure~\ref{fig3}). We measure
a velocity width of $1500\pm300$ km/s using the CIV emission line,
which is lower than the one measured in the SDSS composite QSO
spectrum ($\sim2800$ km/s). On the other
hand, the CIV rest-frame equivalent width ($\rm{EW} = 43\pm10$ {\AA})
is larger than the SDSS QSO composite spectrum ($\sim24$ {\AA}) as
expected from the Baldwin effect, an anti-correlation between 1450
{\AA} luminosity and CIV EW \citep{baldwin77}.  The spectrum also
presents a strong absorption feature between the Ly{$\alpha$} and NV
emission lines. This feature is at approximately the same wavelength
as the [OI] 6364 {\AA} sky line and therefore its real strength is
difficult to estimate. We are nevertheless confident that the
absorption is real and not an artifact of the sky subtraction. We
interpret this absorption line as intrinsic absorption of NV. It is
approximately at 3000 km/s blue-ward of the emission line. Although
the spectrum is noisy, one may tentatively argue that there are also
absorption lines blue-ward of CIV and OVI at approximately the same
velocity offset. If this were correct, that would imply the existence
of material that could obscure the QSO. Another interpretation would
be that the Ly{$\alpha$} emission line is narrow and that we are
seeing the trough between emission lines. Although the wide red tail
of NV would disfavor this hypothesis. The Ly$\alpha$ flux deficit,
$D_A$, is a measure of the absorbed flux blue-ward of Ly$\alpha$
\citep{OK82,schneider91}.We obtain a value of $D_A\sim0.46$ for CXOCY
J125304.0-090737, which is typical of quasars at this redshift (e.g.,
\citealt{madau95}). Our measurement is only approximate due to the
difficulty of fitting the continuum red-ward of Ly$\alpha$ in our poor
signal-to-noise spectrum, which translates into a large uncertainty in
the extrapolation of the continuum blue-wards to compute $D_A$.

Using our photometry, we measure a $V-I=1.13$ color for CXOCY
J125304.0-090737. This value is slightly bluer than the expected
$V-I=1.36$ color obtained redshifting the SDSS composite spectrum
\citep{vandenberk01} and convolving the spectrum with the filter
responses. The similarity between these values argues against large
amounts of intrinsic reddening (and thus optical extinction) in CXOCY
J125304.0-090737. However, this argument is only qualitative and does
not exclude small amounts of extinction.  Overall, considering the
possible relatively weak intrinsic absorption lines in our spectrum of
CXOCY J125304.0-090737 and its not very red photometric properties, we
conclude that there may be small amounts of extinction in CXOCY
J125304.0-090737.



\begin{figure}[th]
\centerline{\psfig{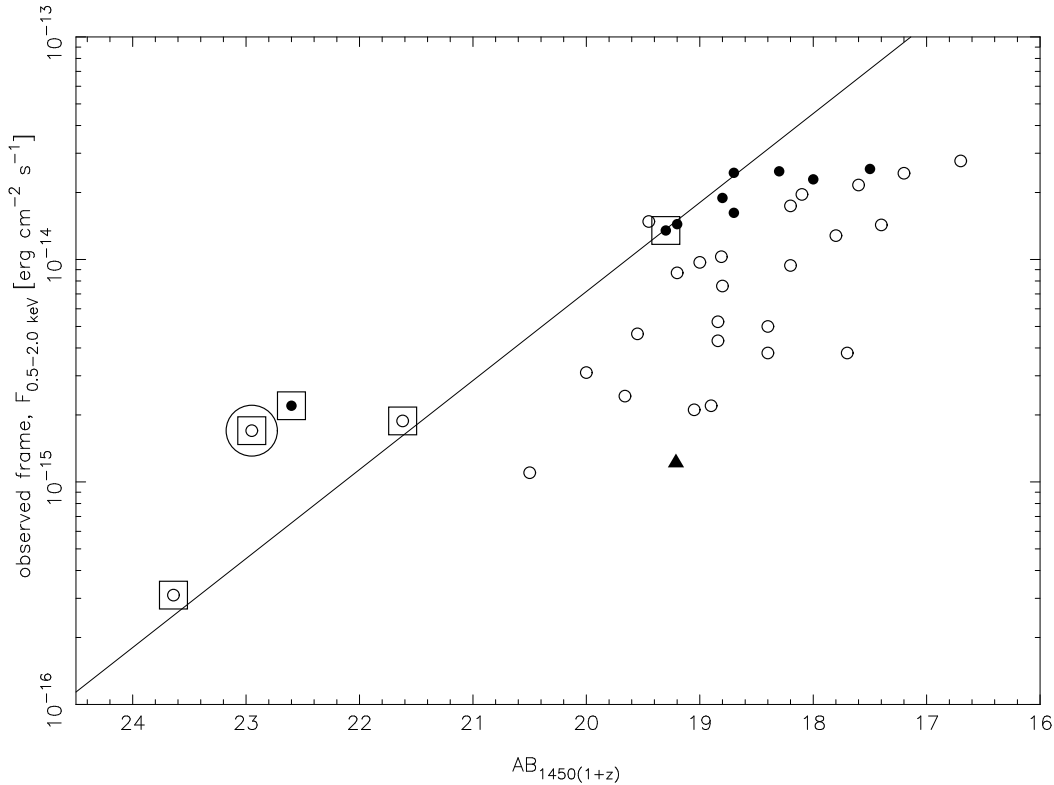}}
\caption{Soft X-ray flux (0.5-2 keV) versus $AB_{1450(1+z)}$. Filled
circles are ROSAT detections, empty circles are Chandra detections and
the filled triangle is the only XMM-Newton detection. Squares enclosing
circles represent X-ray selection, while no enclosing sign denotes
optical selection. CXOCY J125304.0-090737 is emphasized with another
large concentric empty circle at $AB_{1450(1+z)}= 22.95$. Note that
there are no optically selected QSOs detected in X-rays fainter than
$AB_{1450(1+z)}\sim 20.5$. The solid line represents the locus of
objects with $\alpha_{ox}=-1.5$ at $z=4.5$ (see
Fig~\ref{fig5}).\label{fig4}}
\end{figure}

In X-rays, CXOCY J125304.0-090737 is also a faint QSO. The number of
detected photon is low, $17.1\pm4.2$ in the soft [0.5-2.0] keV band
and $8.3\pm3.0$ in the hard [2.0-8.0] keV band, and therefore fitting
a power-law spectrum to the observed counts is difficult (see
Figure~\ref{fig1}). Nevertheless, the best photon index slope is
$\Gamma = 1.69^{+.40}_{-.35}$, which is harder than the average
radio-quiet QSO X-ray spectrum at low redshift
\citep{yuan98,george00,mineo00,RT00} (although part of the difference
may be due to selection biases) and similar to the value of the X-ray
background spectrum, $\Gamma\sim1.4$ (e.g., \citealt{gendreau95};
\citealt{gilli01}). We have used this value ($\Gamma=1.69$) to convert
from observed count rate to X-ray flux. We measure an extinction
corrected X-ray flux of $4.9\pm0.5\times 10^{-15}$ erg cm$^{-2}$
s$^{-1}$ in the observed frame [0.5-8.0] keV band. This QSO is
therefore the second faintest X-ray selected quasar detected so far
(see Figure~\ref{fig4}) and the fourth X-ray faintest high redshift
quasar overall.
 
For spectra with low number of counts it is customary
to characterize the spectral shape with the hardness ratio
$HR=(H-S)/(H+S)$,
where $S$ are the counts measured in a soft band and $H$ are the
counts in a hard band. The measured hardness ratio for CXOCY
J125304.0-090737 is $-0.35^{+0.28}_{-0.25}$, defining the soft and
hard X-ray bands as 0.5-2.0 keV and 2.0-8.0 keV respectively. Although
not many high redshift quasars have measured hardness ratios in the
literature, the ones measured are typically softer than ours. That is
their HR are lower and their $\Gamma$'s are
higher. \cite{vignali02b} measure a typical value of
$\Gamma \simeq 2.0 \pm 0.2$ compared to our $\Gamma\sim 1.7$,
emphasizing the relative hardness of CXOCY J125304.0-090737 compared
to other optically selected high redshift QSOs.

At redshifts $z>4$, the observed Chandra band 0.5-8.0 keV corresponds
to a rest frame band of 2.5-40 keV or higher. Any process changing the
observed spectral slope is thus occurring at intrinsic higher energies
than observed. Strong intrinsic absorption can certainly harden an
observed X-ray spectrum. At redshift $z>4$, Hydrogen column densities
of $N_H \gtrsim 10^{23}$ are needed to produce noticeable spectral
changes. In the case of CXOCY J125304.0-090737 neither the X-ray
spectrum (Fig~\ref{fig1}) nor the optical one (Fig~\ref{fig3}) show
any signs of such amounts of absorption. Processes, such as Compton
reflection (e.g., \citealt{GR88}) which cause deviations from a power
law spectrum at high energies may therefore be important for
understanding the X-ray spectral properties of high redshift quasars
with hard spectra.  The most prominent such features are iron
emission lines, typically seen at 6.4 keV (from a neutral accretion
disk) or 6.7 keV (from an ionized disk), iron absorption edges seen
between 7.0 and 9.0 keV (with the energy again depending on the level
of ionization in the disk), and a broad bump typically between 30 and
50 keV due to the Compton down-scattering of the hight energy photons.
For typical parameters, the spectrum of a z=4.19 quasar can be
hardened by 0.10-0.15 in $\Gamma$ with respect to the underlying power
law by the reflection component.  The effects of reflection components
on the X-ray background have been studied in more detail in past work
(e.g., \citealt{fabian90}; \citealt{zdziarski93}).

In order to compare the optical and X-ray emission we compute the
effective optical-to-X-ray power law spectral slope, $\alpha_{ox}$,
defined as
\begin{equation}
\alpha_{ox}=\frac{\log [f_{\nu}(2\;{\rm keV})/f_{\nu}(2500\;{\rm
\AA})]}{\log[\nu(2\;{\rm keV})/\nu(2500\;{\rm \AA})]}
\end{equation}
where $f_{\nu}$ is the flux density and $\nu$ is the frequency of the
given wavelength or energy. Neither flux density is measured at the
given energy and wavelength and extrapolations are needed to compute
this quantity. For CXOCY J125304.0-090737 we compute
$\alpha_{ox}=-1.35^{+0.11}_{-0.13}$ taking into account the error in
the measured quantities and the errors in the extrapolations.

\begin{figure}[t]
\centerline{\psfig{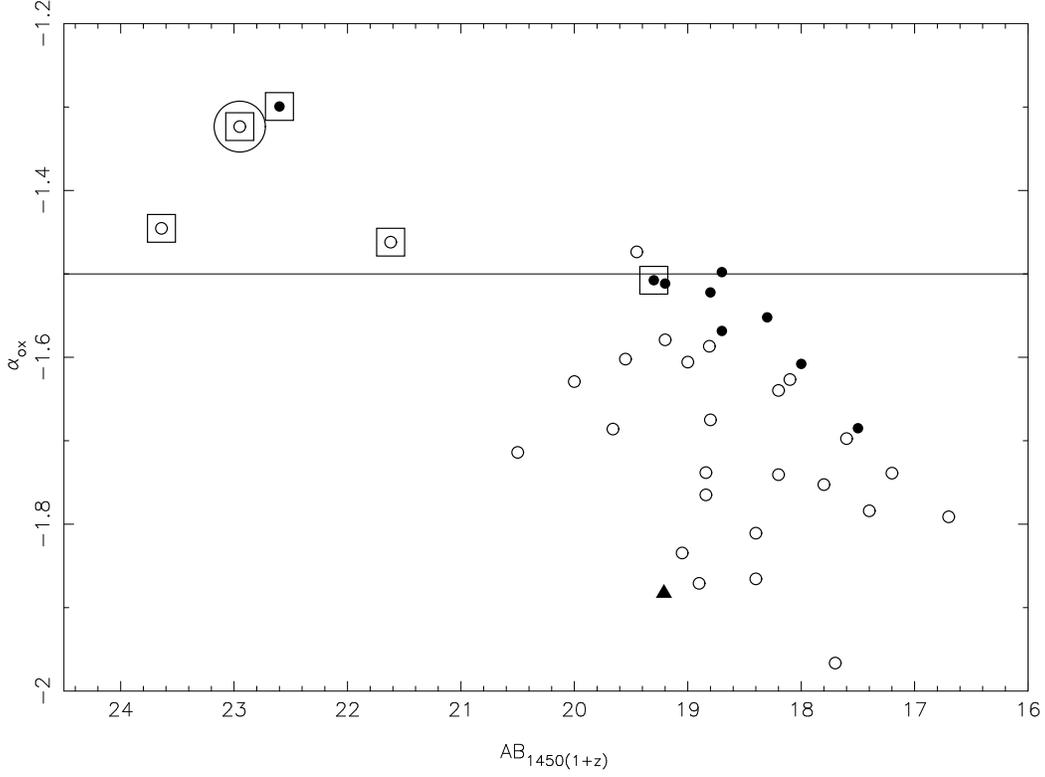}}
\caption{$\alpha_{OX}$ versus $AB_{1450(1+z)}$. Symbols are as in
Fig~\ref{fig4}: 1) small symbols denote detection: ROSAT, filled
circle; Chandra, empty circle; XMM-Newton, filled triangle; 2)
large symbols denote selection: X-ray, square, optical, none. CXOCY
J125304.0-090737 is highlighted with an extra large empty circle at
$AB_{1450(1+z)}= 22.95$. The value of $\alpha_{ox}=-1.5$ is shown with
a straight line. All optically selected QSOs have $\alpha_{ox}$ values
lower than -1.46.\label{fig5}}
\end{figure}

\begin{figure}[t]
\centerline{\psfig{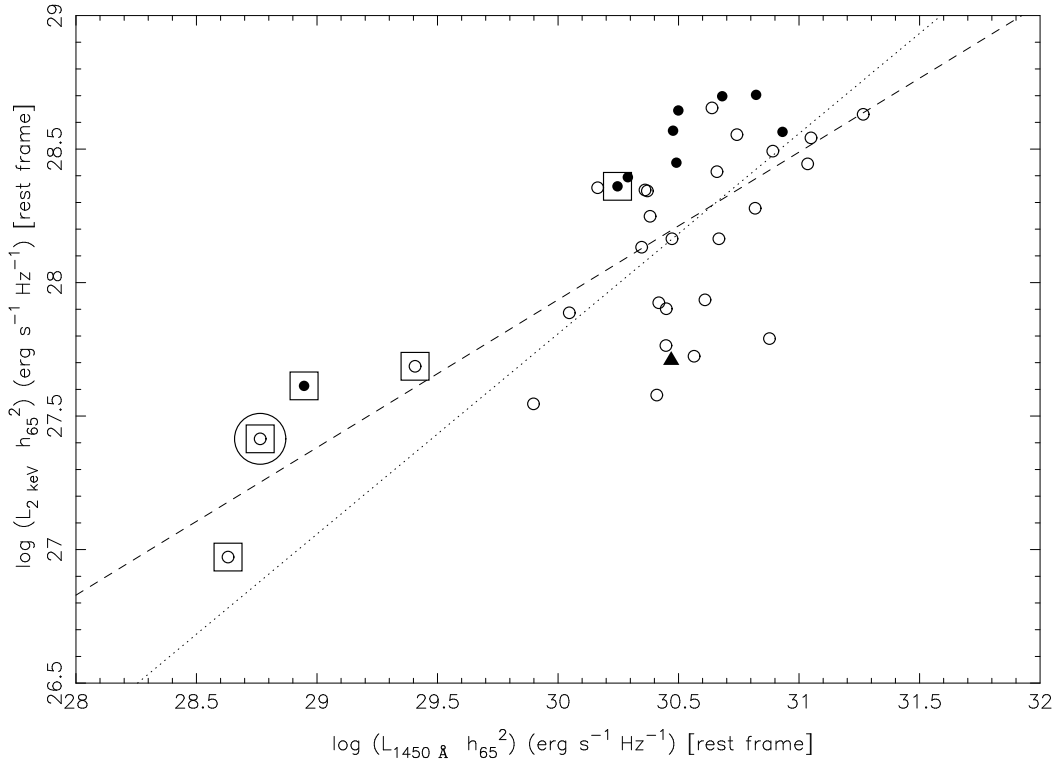}}
\caption{X-ray .vs. optical luminosity densities. Symbols are the same
as in Figures~\ref{fig4} and~\ref{fig5}. The dashed line is the best
fitting linear regression including all points, which gives a slope of
0.55. The dotted line is a fit to the optical selected QSOs fixing the
slope at 0.75 as found by \cite{vignali02d}.\label{fig6}}
\end{figure}

We have also computed $\alpha_{ox}$ for all high redshift quasars
($z>4$) with measured $AB_{1450(1+z)}$ optical magnitudes and
observed-frame 0.5-2.0 keV X-ray fluxes available in the literature.
Figure~\ref{fig5} plots $\alpha_{ox}$ against the optical magnitude
$AB_{1450(1+z)}$. The $\alpha_{ox}$ values of the X-ray selected QSOs
are shallower (X-ray stronger and/or optically fainter) than the
optically selected ones. A Kolmogorov-Smirnov (KS) test indicates that
the probability of both distribution being drawn from the same parent
distribution is less than 0.05\%. The distribution of optical absolute
magnitudes (or luminosities) is also statistically different in the
X-ray and optically selected samples because the X-ray surveys are
more sensitive but cover a much smaller area than the optical ones
(see Figures~\ref{fig4} and~\ref{fig6}). In order to understand the
difference in the $\alpha_{ox}$ values, we plot the X-ray against the
rest frame UV luminosity densities in Figure~\ref{fig6}. We find that
both observables are correlated with a functional form $L_{2\;keV}
\propto L_{1450\;{\AA}}^{0.55}$, shown as a dashed line in
Figure~\ref{fig6}.  This slope is somewhat shallower than that found
by \cite{vignali02d}.  But taking into account that the faint UV
luminosity points, driving a shallower slope, are X-ray selected and
assuming that the distribution of X-ray luminosities are the same at
faint UV rest frame luminosities than at bright UV luminosities, our
data are consistent with the correlation found by
\cite{vignali02d}, thus extending their result to fainter
luminosities at high redshift with a larger sample.

CXOCY J125304.0-090737 was discovered in the first Chandra X-ray
Southern field studied within the CYDER survey. One can ask how many
high redshift QSOs one would expect to detect given the QSO luminosity
function. We use \citet{fan01a} QSO optical luminosity function (LF),
which is consistent with previous estimates (e.g.,
\citealt{kennefick95}; \citealt{schmidt95}).  They compute the
optically selected QSO LF in the range $3.6<z<5.0$ and
$-27.75<M_{1450}-5\times log (h_{65})<-25.75$. The optical absolute
magnitude of CXOCY J125304.0-090737 is considerably fainter. If we
extrapolate Fan et al's power law LF down to the magnitudes reached by
our search, the expected number of quasars in the ACIS-S S3 CCD with
$z>4$ is 0.22. The probability of finding {\it any} quasars (one or
more) is then 20\%. If instead we extrapolate Fan et al's LF to
fainter luminosities using the shape of the optical QSO LF at lower
redshifts \citep{boyle00}, the expected number of QSOs is even lower.
Although, with just one detection these numbers are not significant,
they are nevertheless indicative that the faint end slope of the QSO
optical LF may be high (at least higher than at low redshift) or that
optical surveys miss significant fractions of QSOs that may be
obscured. In the case of CXOCY J125304.0-090737 is worth pointing out
that its colors are typical of low redshift quasars and therefore a
multi-wavelength optical survey reaching this magnitude limit should
have been able to pick it up.

The number of expected $z>4$ quasars can also be computed from the
X-ray luminosity function (XLF). Unfortunately, the XLF is poorly
determined above $z=4$, given the lack of X-ray selected QSOs at these
redshifts. Compiling several X-ray surveys, \cite{miyaji01} derive an
XLF up to $z=4.6$, although at their highest redshift their XLF is
only valid at the bright end. If we extrapolate their power-law
parameterization of their last redshift bin ($z=2.3-4.6$) to the faint
end and also assume that the XLF is the same beyond $z=4.6$, then we
would expect to detect 0.72 quasars at the X-ray flux limit reached in
our first CYDER field. If we instead extrapolate Miyaji et al's XLF
with a smooth double power-law to the faint end similar to the one at
lower redshifts and also extrapolate this function to higher redshift
with a plausible evolutionary model (see Castander et al. in
preparation for details), we obtain somewhat lower expectation
values. The probability of finding any quasars would then be $\lesssim
50\%$. We want to stress nevertheless that these numbers should be
taken cautiously. They are only orientative to give a sense of how the
faint end of the high redshift quasar luminosity function may behave.

The discovery of CXOCY J125304.0-090737, together with the other X-ray
selected high redshift QSOs
\citep{henry94,zickgraf97,schneider98,silverman02,barger02},
demonstrates the potential of X-ray techniques to detect high redshift
QSOs. At redshifts $z>4$, X-ray photons observed for example at 1 keV
were emitted at energies $>5$ keV, where the X-ray photons can
penetrate even considerable amounts of extinction and therefore X-ray
selected samples should be almost extinction-bias free. Only sources
with intrinsic column densities $N_H \gtrsim 10^{23}$ cm$^2$ should be
affected. In particular, Chandra with its superior image quality that
allows secure and economical identifications of optical counterparts
will permit the completion of follow-up surveys, some of them which
are already in progress, like the CYDER survey, that augur a new
understanding of the quasar population at high redshift.  For example,
one can argue that one needs of the order of 100 high redshift quasars
to constraint the high redshift LF. This could be achieved with 40-200
Chandra pointings (ACIS-I and ACIS-S) depending on whether the actual
QSOs LF at the faint end is as our discovery implies or
lower. Although demanding, current on-going X-ray follow up surveys
may achieve these numbers. These discoveries will be complemented by
deep wide-field multi-band optical surveys to provide a good
understanding of the high redshift quasar population and thus the
history of the accretion power onto black holes which is thought to be
closely related to the galaxy formation and star formation rate
history of the universe.

\section{Conclusions}

We present the discovery of CXOCY J125304.0-090737, a new high
redshift quasar at $z=4.179$ found as part of the CYDER survey. CXOCY
J125304.0-090737 is an optically faint quasar ($M_B=-23.69+5\times log
(h_{65})$) with a typical QSO spectrum (Figure~\ref{fig3}). In the
optical, the quasar seems to be slightly obscured, although the data
is also consistent with no extinction. In X-rays, CXOCY
J125304.0-090737 is also X-ray faint ($f_X = 1.7\pm 0.4 \times
10^{-15}$ ergs s$^{-1}$ cm$^{-2}$ in the [0.5-2.0] keV band) with a
somewhat harder spectrum ($\Gamma\sim1.7$ or $HR\sim-0.35$) than
typical low redshift or high redshift optically selected quasars. We
speculate that a reflection component can slightly harden the spectrum
but by no means is this the only mechanism. The spectrum of CXOCY
J125304.0-090737 and the other X-ray selected high redshift quasars is
X-ray strong (and/or optically weak) compared to high redshift
optically selected quasars. Given that both samples have different
optical/ultraviolet absolute magnitudes, this result is a
manifestation of the correlation between X-ray and rest frame
ultraviolet luminosities. Taking into account selection biases, we
find a slope for this correlation consistent with the value obtained
by \cite{vignali02d}. We therefore extend their result to lower
luminosities at high redshifts using a larger sample.

The quasar space density implied by the discovery of CXOCY
J125304.0-090737 is high: a factor $\gtrsim5$ larger than reasonable
extrapolations of the quasar optical luminosity function at high
redshift. Although not statistically significant, this number seems to
indicate that the shape of the quasar luminosity is changing with the
faint end slope becoming steeper with redshift or that current optical
surveys at high redshift are missing considerable amounts of
quasars. The space density predicted from the X-ray luminosity
function is more a guess than a prediction. Anyway, if on-going and
future X-ray surveys corroborate the numbers implied by the first
field of the CYDER survey, the X-ray luminosity function will not
differ much of its value at lower redshifts, $z=2-3$. The combination
of the currently ongoing optical and X-ray surveys with complementary
regions of the parameter space sampled will help to determine the
quasar luminosity function at high redshift with a better
understanding of the biases involved with both selections. The CYDER
survey promises to be one of the contributors to a better
understanding of the quasar population at high redshift.

The CYDER survey participants and especially FJC and ET acknowledge
support from the Fundaci\'on Andes. JM acknowledges support from
FONDECYT.

\begin{deluxetable}{cc}
\tablecaption{Summary of observed properties of CXOCY J125304.0-090737\label{tab1}}
\tablecolumns{2}
\tablewidth{0pt}
\startdata
\tableline\tableline
Parameter & Value \\
\tableline
RA (J2000) & $12^h53^m04.0^s$ \\
Dec (J2000) & $-09^o07'37''$ \\
N$_H$\tablenotemark{a} & $2.96\times 10^{20}$ cm$^{-2}$\\ 
$z$ & $4.179\pm0.006$ \\
$\alpha_{OX}$ & $\alpha_{ox}=-1.35^{+0.11}_{-0.13}$\\ 
V mag (Vega) & $23.65\pm 0.05$ \\
I mag (Vega) & $22.52\pm 0.08$ \\
J mag (Vega) & $21.9\pm 0.2$ \\
K$_s$ mag (Vega) & $>20.5$ \\
$M_I$ & $-23.45 + 5\times log (h_{65})$\\
$AB_{1450(1+z)}$ & $22.95\pm0.10$\\
$M_{1450(1+z)}$ & $-23.0 + 5\times log (h_{65})$\\
$f_X$ (0.5-8 keV) & $4.9\pm 1.3 \times 10^{-15}$ ergs s$^{-1}$ cm$^{-2}$\\
$f_X$ (0.5-2 keV) & $1.7\pm 0.4 \times 10^{-15}$ ergs s$^{-1}$ cm$^{-2}$\\
$f_X$ (2.0-8 keV) & $3.2\pm 1.0 \times 10^{-15}$ ergs s$^{-1}$ cm$^{-2}$\\
Hardness Ratio & $-0.35^{+0.28}_{-0.25}$ \\
\enddata
\tablenotetext{a}{calculated using HEASARC tool nh}
\end{deluxetable}


\begin{thebibliography}{}
\bibitem[Anderson et al.(2001)]{anderson01} Anderson, S. F., et al., 2001, \aj , 122, 503
\bibitem[Baldwin et al.(1977)]{baldwin77} Baldwin, J. A., 1977, \apj, 214, 679
\bibitem[Barger et al.(2002)]{barger02} Barger, A. J., et al., 2002, \aj, 124, 1839
\bibitem[Bechtold et al.(2002)]{bechtold02} Bechtold, J., et al., 2002, \apj, in press, astro-ph/0204462
\bibitem[Bennett(1962)]{bennett62} Bennett, A. S., 1962, Mem. R. Astron. Soc., 68, 163
\bibitem[Boyle et al.(2000)]{boyle00} Boyle, B. J., et al., 2000,
\mnras, 317, 1014
\bibitem[Brandt et al.(2001)]{brandt01} Brandt W. N., et al., 2001, \aj, 122, 1
\bibitem[Brandt et al.(2002a)]{brandt02a} Brandt W. N., Schneider
D. P., et al., 2002a, \apj, 569, L5
\bibitem[Brandt et al.(2002b)]{brandt02b} Brandt W. N., Schneider
D. P., et al., 2002b, in X-ray Spectroscopy of AGN with Chandra and XMM-Newton, eds, Boller et al, (Garching: MPE press), 235
\bibitem[Castander et al.(2003)]{castander03} Castander, F. J., Treister, E., Maza, J., Coppi, P., Maccarone, T., Zepf, S., Guzm\'an, R. \& Ruiz, M. T., 2003, AN, in press, astro-ph/0211643
\bibitem[Edge et al.(1959)]{edge59} Edge, D. O., Shakeshaft, J. R., McAdam, W. B., Baldwin, J. E. \& Archer, S., 1959, Mem. R. Astron. Soc., 68, 37
\bibitem[Fabian et al.(1990)]{fabian90} Fabian, A. C., et al., 1990, \mnras, 242, P14
\bibitem[Fan et al.(2001a)]{fan01a} Fan X., et al., 2001b, \aj ,121, 54
\bibitem[Fan et al.(2001b)]{fan01b} Fan X., et al., 2001c, \aj ,122, 2833
\bibitem[Gendreau et al.(1995)]{gendreau95} Gendreau, K. C., et al, 1995, PASJ, 47, L5
\bibitem[George et al.(2000)]{george00} George, I. M., et al., 2000, \apj, 531, 52
\bibitem[Gilli et al.(2001)]{gilli01} Gilli, R., Salvati, M. \& Hasinger, G., 2001, A\&A, 366, 407
\bibitem[Guilbert \& Rees(1988)]{GR88} Guilbert, P. W. \& Rees, M. J., 1988, \mnras, 233 475
\bibitem[Henry et al.(1994)]{henry94} Henry, P.J., et al., 1994, \aj , 107, 1270
\bibitem[Kaspi et al.(2000)]{kaspi00} Kaspi S., Brandt, W. N. \& Schneider, D. P., 2000, \aj , 119, 2031
\bibitem[Kennefick et al.(1995)]{kennefick95} Kennefick, J. D., Djorgovski, S. G. \& de Carvalho, R. R., 1995, \aj, 110, 2553
\bibitem[Madau(1995)]{madau95} Madau, P., 1995, \apj, 441, 18
\bibitem[Mineo et al.(2000)]{mineo00} Mineo, T., et al., 2000, A\&A, 359, 471
\bibitem[Miyaji et al.(2001)]{miyaji01} Miyaji, T., Hasinger, G. \& Schmidt, M., 2001, A\&A, 369, 49
\bibitem[Oke \& Korycansky(1982)]{OK82} Oke, J. B. \& Korykansky, D.G., 1982, \apj, 255, 11
\bibitem[Reeves \& Turner(2000)]{RT00} Reeves, J. N. \& Turner, M. J. L., 2000, MNRAS, 316, 234
\bibitem[Richards et al.(2003)]{richards03} Richards, G. T., et al., 2003, AJ, submitted
\bibitem[Sandage(1965)]{sandage65} Sandage, A., 1965, \apj, 141, 1560
\bibitem[Schneider et al.(1991)]{schneider91} Schneider D.P, Schmidt, M. \& Gunn, J. E.,  1991, \aj , 101, 2004
\bibitem[Schneider et al.(1998)]{schneider98} Schneider D.P, et al., 1998, \aj , 115, 1230
\bibitem[Schneider et al.(2002)]{schneider02} Schneider D.P, et al., 2002, \aj , 123, 567
\bibitem[Silverman et al.(2002)]{silverman02} Silverman, J. D., et al., 2002, \apj , 569, L1
\bibitem[Schmidt(1963)]{schmidt63} Schmidt, M., 1963, Nature, 197, 1040
\bibitem[Schmidt et al.(1995)]{schmidt95} Schmidt, M., Schneider, D. P. \& Gunn, J. E., 1995, \aj, 110, 68
\bibitem[Treister \& Castander(2003)]{treister03} Treister, E. \& Castander, F. J., 2003, AN, in press. 
\bibitem[Vanden Berk et al.(2001)]{vandenberk01} Vanden Berk, D.E.,
et al., 2001, \aj, 122, 549
\bibitem[Vignali et al.(2001)]{vignali01} Vignali, C., et al., 2001, \aj, 122, 2143
\bibitem[Vignali et al.(2002a)]{vignali02a} Vignali, C., et al., 2002a, in proceedings of the 5th Italian AGN Meeting "Inflows, Outflows and Reprocessing around black holes", astro-ph/0210001
\bibitem[Vignali et al.(2002b)]{vignali02b} Vignali, C., et al., 2002b, \aj, in press, astro-ph/0210475
\bibitem[Vignali et al.(2002c)]{vignali02c} Vignali, C., et al., 2000c, \apj, in press, astro-ph/0210552
\bibitem[Vignali et al.(2002d)]{vignali02d} Vignali, C., Brandt, W. N. \& Schneider, D. P., 2000d, \aj, in press, (astro-ph/0211125)
\bibitem[White et al.(2000)]{white00} White, R.L., et al, 2000, \apjs, 126, 133
\bibitem[York et al.(2000)]{york00} York, D. G., et al., 2000, \aj, 120, 1579
\bibitem[Yuan et al.(1998)]{yuan98} Yuan, W., Brinkmann, W., Siefert, J. \& Voges, W., 1998, A\&A, 330, 108
\bibitem[Zdziarski et al.(1993)]{zdziarski93} Zdziarski, A.A., et al., 1993, \apj, 405, 125
\bibitem[Zickgraf et al.(1997)]{zickgraf97} Zickgraf, F.-J., et al., 1997, A\&A, 323, L21
\end{thebibliography}
\end{document}